\documentstyle[11pt]{article}

\begin{document}

\title{On the entropy devil's staircase in a
                        family of gap-tent maps }

\author{Karol {\.Z}yczkowski$^*$
 \\
Center for Plasma Research\\ University of Maryland\\
College Park, MD 20742\\
\and
Erik M. Bollt\\
Mathematics Department\\
572 Holloway Rd \\
U.S. Naval Academy \\
Annapolis, MD 21402-5002.}
\maketitle

\vspace{0.6cm}
\begin{abstract}

To analyze the trade-off between channel
capacity and noise-resistance in designing dynamical systems to pursue
the idea of
communications with chaos, we perform a measure theoretic analysis the
topological entropy function of a ``gap-tent map" whose
invariant set is an unstable chaotic saddle of the tent map.  
Our model system, the ``gap-tent map" is a family of tent
 maps with a symmetric gap, which mimics the presence
of noise in physical realizations of chaotic systems, and for this
 model, we can perform many calculations in closed form.
We demonstrate that the dependence of the topological entropy on the
size of the gap has a structure of the devil's staircase. 
 By integrating over a fractal measure, we obtain
analytical, piece-wise differentiable approximations of this
dependence.  Applying concepts of the kneading theory we find the
position and the values of the entropy for all leading entropy
plateaus. Similar properties hold also for the dependence of the
fractal dimension of the invariant set and the escape rate.

\end{abstract}

\vspace{1.2cm}

\centerline{ PACS: {\it 05.45.+b, 47.52.+j, 47.53.+n, 95.10.Fh}}
\vspace{0.8cm}


\vspace{0.8cm}
\centerline{{\small{
e-mail: $^1$karol@chaos.if.uj.edu.pl
\ \ \ \ \ \ \ \ \ $^2$bollt@nadn.navy.mil  }}}

\vfil
$^*$ Permanent address: Instytut
Fizyki im. Smoluchowskiego, Uniwersytet Jagiello{\'n}ski,
ul. Reymonta 4, 30-059 Krak{\'o}w, Poland 
\newpage

\setlength{\baselineskip}{1.0\baselineskip}
\renewcommand{\baselinestretch}{2}  

\section{Introduction}

Long term evolution of a chaotic system is, by definition,
unpredictable.
The topological complexity of a dynamical system has classically been
measured in terms of {\it topological entropy} \cite{AKM:1965}, which
may be
considered as the growth rate of distinct states of the dynamical
system, to
the myopic observer. Shannon's information theory tells us that a
sequence of
events conveys information only if the events are not fully predictable,
and
therefore, the topological entropy may be considered as a quantitative
measurement of the information generating capacity of the chaotic
dynamical
system
\cite{Shannon:1948,Blahut:1988}.  On the other hand, metric-entropy can
also
be considered as a statistical quantification of irregular behavior,
where
probability of events are weighted in terms of the  specific invariant
measure chosen. Indeed, the link between these two types of
quantifiers, topological and metric entropy, is in terms of the choice of
invariant measure.

Given this classical and deeply rooted link between chaos and
information
theory, perhaps it is surprising that only recently we have finally
realized
that the link can be exploited, since chaotic oscillators can be
controlled
to transmit messages, \cite{Hayes,cuomo}.  However, such exploitation of
chaotic oscillators awaited the (now obvious) realization that chaos is
controllable; following the seminal ``OGY" paper \cite{OGY}
at the beginning of this decade, there has literally been an explosion
of
various approaches and wide applications of the concept.  The oxymoron
between the words, ``chaos" and ``control," is resolved by the fact that
a
chaotic dynamical system is deterministic, even while it is long-term
unpredictable.

The Hayes et.~al.~\cite{Hayes} method of chaos communication relies on
the
now classical description of a chaotic dynamical system in terms of a
(semi)-conjugacy (an equivalence) to the Bernoulli-shift map symbolic
dynamics \cite{Smale,Robinson}.
Given the link of a conjugacy function, i.e., a coding function, control
of
chaotic trajectories is equivalent to control of (message bearing)
digital
bits.
However, there were practical obstacles: 1) experimentally observed
chaotic
dynamical systems are typically nonhyperbolic
\cite{lai-dong-grebogi-yorke-kan} and therefore Markov
 generating partitions
are difficult to specify \cite{GK:1989,Cvit:1988}, 2) even small channel
noise could cause bit-errors.  For example, in the case of a two-symbol
dynamics representation, a point {\bf x} near the symbol partition,
which
bears the message bit, say a {\bf 0}, may be kicked across the partition
by
external noise, and therefore an the error-bit, a {\bf 1},
 is inadvertently
transmitted.  One of us has recently co-authored a technique to solve
both of
these problems \cite{bollt-lai}.  Avoiding neighborhoods of the
generating
partition,  and {\it all pre-iterates} of the partition, yields an
invariant
Cantor-like  unstable chaotic saddle, which we showed, is robust to
reasonably high noise amplitudes.  Interestingly, the topological
entropy of
these unstable saddles was found to be a monotone nonincreasing devil's
staircase-like function of the noise gap width $s$.  In a subsequent
work
\cite{bollt-lai2}, this devil's staircase function was analyzed from a
topological standpoint, by a ``bifurcation" diagram in $s$, of the
word-bins,
which were found to collide with each other at varying $s$ values, thus
creating the ``flat-spots."
While we have initially been motivated to study the one-d maps with a
gap,
due to the communications application, we have subsequently also found
the
analysis of their ergodic properties to be quite rich.

In this paper, we shall be concerned with ergodic properties of the {\sl
tent with a
gap} - a 1D dynamical system defined for
$x\in Y=[0,1]$ by \cite{bollt-lai}
\begin{equation}
 f_{\varepsilon}(x)=  \left\{ \begin{array}{lll}
 2x  &~\hbox{for}~&  x \le  (1-\varepsilon)/2 \nonumber \\
 -1 ~ ~ \hbox{(gap)}~ &~\hbox{for}~&  x
  \in \bigl( (1-\varepsilon)/2, (1+\varepsilon)/2\bigr) \\
 2(1-x)  &~\hbox{for}~&  x \ge  (1+\varepsilon)/2 \nonumber \\
\end{array}
\right.
\label{ten}
\end{equation}
where the size of the gap $\varepsilon $ is a free parameter
$0\le\varepsilon < 1$.   The value of the function inside the gap $(-1)$
is an arbitrary number outside the interval $[0,1]$.  See Fig.~1.  This
map
has the desired property that orbits have reduced probability of bit
error
due to noise of amplitude less than $\varepsilon$ \cite{bollt-lai}.
For
simplified analysis in this paper, we consider these flattened tent maps
Eq.~(\ref{ten}), for which many calculations are in closed form, and
which
may be considered to be typical of truncated continuous one-hump maps.
 The entropy of the
 closely related trapezoidal maps (flat ``roof" of the tent instead of
the
gap) was studied in \cite{BMT91}, but their analysis was quite
different,
involving splitting the family of maps into a codimension-one foliation
according to possible dynamical behaviors.

The kneading theory of Thurston and Milnor, introduced in
Ref.~\cite{MT77}, revealed the importance of the itineraries of
the critical points, the so-called {\it kneading sequences}, for the
admissibility of other itineraries. Kneading theory gives a bifurcation
theory of symbol sequences, which for one parameter ``full families" of
maps,
including the logistic map $x_{n+1}=\lambda x_n(1-x_n)$ and the tent map
$x_{n+1}=s(1/2-|x-1/2|)$.
The symbol dynamical representation of
bifurcations includes the periodic windows in Sharkovsky's order, and
corresponding flat spots on the topological entropy versus parameter
value
function.  ``Fully developed chaos," occurs when these maps are
everywhere
two-to-one ($\lambda>4$, or $s>2$); there is a conjugacy to the
fullshift
and
the topological entropy is $h_T=ln(2)$.  Correspondingly, the
topological
entropy is found to be constant in periodic windows
\cite{MT77,ER85,CCP95a,HO97},
which are known to be dense in the parameter $\lambda$ for the logistic
map, and a closely related, quadratic map \cite{GS97}.  As we shall
clarify in this paper,
an analogous situation occurs for our model noise-resistant tent gap-map
Eq.~(\ref{ten}); as the gap parameter $\varepsilon$ is increased the
maximal
value of the map is decreased, and correspondingly the kneading sequence
is
decreased, with now familiar consequences to the topological entropy
function, which also has flat spots in the window regions.  In fact,
staircase like functions can be found repeatedly in the dynamical
systems literature \cite{devilish}, and they often signify structural
stability of the observed quantity.

A brief overview of this paper is as follows.  In Section 2, we will
make the necessary introductory definitions and standardize notation,
concerning entropy, as well as gap boundaries, and preimages, and
relevant measures.  In Section 3, we derive conceptually simple and new
formulas for the direct computation of the devil's staircase topological
entropy function, summarized by Proposition 1.  In Section 4, we briefly
demonstrate, by example, the efficacy of the formula of Proposition 1.
Then in Section 5, we discuss implications to the structure of the
devil's staircase function, which we can conclude using the formula, and
the closed form representation of the endpoints of the ``flat-spots."
In Section 6, (and Appendix A), we link the roots of the corresponding
polynomials to the topological entropy (and we relate this to the
kneading theory of Milnor and Thurston \cite{MT77}).  We show how the
infinitely fine structure of the devil's staircase function is
correspondingly linked to a sequence of ``descendant" polynomials, by
several operators on sequence space, which we shall discuss.
Furthermore, we discuss the structure of the flat-spots, in Proposition
2, and we relate this to the familiar Feigenbaum-like accumulation.
This leads to several exact statements which can made concerning values
of the entropy at specific points, as well as a sharp estimate for the
``maximal" topologically nondegenerate gap, $e_*$.  In Section 7, we
discuss a generalization based on weakening the assumptions of
Proposition 1, yielding Proposition 3 in which we show that topological
entropy of a class of maps can be calculated by ``averaging" (with
respect to the appropriate measure) the pointwise pre-image count.
Finally, in Section 8,  we discuss implications to the pointwise
spectrum of fractal Renyi-dimensions, as well as escape rates.

\section{Measures and Entropies of the Tent Gap-Map}

 Before we may begin our main propositions in the following sections, we
have
several preliminary definitions and derivations to present in this
section.

The topological entropy of a map f on its phase space $Y$, may be
defined in
the symbol space representation as the asymptotic rate of growth of
permissible words with respect to word length,
\begin{equation}\label{topent1}
h_T = \lim_{n\rightarrow\infty}\frac{\ln{N(n)}}{n},
\end{equation}

where $N(n)$ is the number of permissible words of length $n$.
Alternatively, we may define the topological entropy to be
\begin{equation}
h_T=sup_{\mu\in \mbox{{\it M}}_{I}(f)} h(\mu,f),
\end{equation}
where we take the supremum over entropies with respect to the set of all
invariant measures  $\mu\in\mbox{{\it M}}_{I}(f)$, of $f$.  For a
topological
Markov chain, such as $f_\varepsilon$, this supremum is actually
attained by the
entropy with respect to the Parry measure $\mu_*$, $h_T(f)=h(\mu_*,f)$
\cite{Katok}, also known as Parry's topological entropy measure.

The use of formula Eq.~(\ref{topent1}) requires splitting the phase
space
$Y=[0,1]$ into cells of equivalent $m$-iterate $f_\varepsilon$ orbits,
which we
can write in closed form for our model tent gap-map $f_\varepsilon$; we
will use
the formula for the m-bit bins extensively throughout the rest of this
paper.
 In fact, our ability to write the m-bit bins $A_{mk}$ in closed form,
allows
us to explicitly perform many of the calculations in this paper, which
was a
main motivation of choosing the model Eq.~(\ref{ten}).
To find the (up to) $2^m$, $m$-bit bins, $A_{mk}$, consider the critical
point $x_c=1/2$ of the tent map $f_0$. It
has $2^m$ $m$-th preimages
   with respect to the tent map (without the gap). They
   are $f_0^{-m}(x_c)=(2k-1)/2^{m+1}$ with
$k=1,2,\dots,2^m$.
Let $A_{01}$  denote  the gap
$A_{01}= \bigl( (1-\varepsilon)/2, (1+\varepsilon)/2\bigr)$.
      Its $m$-th preimages with respect to the system
      $f_{\varepsilon}$ have the  width $\varepsilon/2^m$ and are
 centered at the preimages of the critical  point.
Denoting $m$--th preimages of the gap by $A_{mk}$ we obtain
\begin{equation}
A_{mk}= \Bigl( {2k-1-\varepsilon \over 2^{m+1}},
{ 2k-1+\varepsilon \over 2^{m+1}} \Bigr),
\label{amk}
\end{equation}
 where  $m=1,2,3,\dots$ and $k=1,\dots,2^m$. We note
that depending on the size of the gap $\varepsilon$
some preimages, corresponding to different values of $m$, may merge
together, and it is exactly such overlaps which cause topological
entropy
flat spots.  Also note that each of the $m$-bit word cells $A_{mk}$ are
the
same length (Lebesgue measured); this is due to the fact that our map
$f_\varepsilon$ has a constant magnitude slope, $|f_\varepsilon'|=2$.

We also find it useful to consider,
 $\mu_{SRB}$, the SRB invariant measure
for the system $f_{\varepsilon}$. It may be defined as
the eigenmeasure of the Koopman operator associated with the
system (i.e. the adjoint
to Frobenuis-Perron operator) corresponding to the largest eigenvalue.
A detailed discussion of these measures can be found in
Ref.~\cite{BR87} for ``cookie  cutters."

\smallskip
{\bf Proposition 0:} {\it The  support of $\mu_{SRB}$ is contained in
$X=[f_{\varepsilon}(y_m),y_m]=[2\varepsilon,1-\varepsilon]$,
where $y_m$  is the maximal value of the system
  $f_{\varepsilon}((1-\varepsilon)/2)=1-\varepsilon$}, (see Fig.~1).

\smallskip
The  support of $\mu_{SRB}$ equals $S=\lim_{n\to \infty} S_n$, where
\begin{equation}
   S_n=X \setminus \bigcup_{m=0}^n f^{-m} (A_{00}) =
   X \setminus \bigcup_{m=0}^n \bigcup_{k=1}^{2^m} A_{mk}
\end{equation}
Let us also define the  {\sl uniform} measure $\mu_u$, constructed
inductively on $S$  as follows:
\begin{eqnarray}
 &\bullet&\mbox{ let }\mu_1\mbox{ be the uniform measure on }S_1\mbox{
     normalized so that  }\int_X d\mu_1=1; \\
\label{uniformn}
 &\bullet&\mbox{ let }\mu_n\mbox{ be the uniform normalized measure on
}S_n;\\
 &\bullet&\mbox{ the uniform measure }\mu_u\mbox{ is defined  by the
weak limit
      }\mu_u=\lim_{n\to \infty} \mu_n.
\end{eqnarray}

The uniform measure $\mu_u$ is connected to the {\sl
natural measure on stable manifold} $\mu_s$, which is often considered
in
the physical literature \cite{Ott94}. Let us draw randomly, $N(0)$
initial
points with respect to the uniform density on
the set $Y=[0,1]$ which contains $S$, and let $N(n)$ denote
the number of points which did not leave the system by the $n$-th
iterate.
Let $C$ be any Borel subset of $Y$, and
$N_s(C,n)$ denote  the number of  initial conditions belonging to
$C$ whose trajectories remain in the system for $n$ iterations. Then
one defines \cite{Ott94}
\begin{equation}
   \mu_s(C) = \lim_{n\to \infty} \lim_{N(0)\to \infty}
      {N_s(C,n) \over N(n)}.
\label{munat}
\end{equation}

\smallskip
 {\bf Remark} for $x\in S$ both measures are proportional,
  $\mu_u(A)= c \mu_s(A)$, for Borel subsets $A\subset Y$, where the
proportionality constant
  $c$ depends on the choice of the set $Y$.

For the case of the tent gap-map $f_\varepsilon$, in Eq~(\ref{ten}), we
can show
that the $\mu_{SRB}$, corresponding to natural measure, coincides with
the
Parry's maximal entropy measure, $\mu_*$.  To show this, we observe that
the
absolute value of the
derivative is constant: $|f'(x)|=2$ for any $x\in Y$.    Hence, the
topological entropy and
the KS metric entropy are equal $h_T=h_{KS}=h_q$.  The
notation
$h_q$
denotes the generalized Renyi entropies, which are discussed in
detail in \cite{Renyi}.  This coincidence follows since the average
 $\int_Y  \ln |f'(x)| d\mu(x)$ equals ln$2$, independently of the
measure $\mu$. Hence, we see that the topological pressure, which is
defined
\cite{BR87},
\begin{equation}
P(\beta)=\sup_{\mu} [h_{KS}(\mu)-  \int_Y  \beta\ln |f'(x)| d\mu(x)],
\label{press}
\end{equation}
where the supremum  is taken over
all invariant measures of $f$, has the  constant integral fixed at
ln$2$, for
all measures $\mu$.  It follows then that supremum is achieved,
by the measure which we denote $\mu_{SRB}$, irrespective of the
parameter $\beta$. Therefore, in this
case, it follows that the SRB measure coincides with the maximal entropy
measure, $\mu_*=\mu_{SRB}$.

\section{Topological entropy expressed by mean number of preimages}

In \cite{bollt-lai,bollt-lai2,JacobsOttHunt}, it was shown that the
topological entropy function $h_t(\varepsilon)$ is a ``devil's
staircase"
function for the model $f_\varepsilon$.  We are now in a position to
better understand the flat spots  of the staircase.  In this section, we
also give a Proposition, which gives  an explicit construction to
generate a sequence of  approximating functions which asymptotically
converge
to the devil's staircase.

We begin by noting that the point $f^2_{\varepsilon}
 (y_m)=4\varepsilon $ divides the support  $S$ into two
 sets: $E_2$ contains all points $x \ge 4 \varepsilon$ of the
support  which  have  two preimages
with respect to $f_{\varepsilon}$, and $E_1=S \setminus E_2$ consists
of points with  one preimage only.  See Fig.~1.

Let us define a function $M(\varepsilon)$
\begin{equation}
  M (\varepsilon) := \int_{4\varepsilon}^{1-\varepsilon} d\mu_u,
\label{mm}
\end{equation}
which is closely related to the topological entropy function.  Notice
that
$M(\varepsilon)$ measures the relative ``volume" of the set with two
preimages,
$\mu_u(E_2)$.

 The  right hand side of (\ref{mm}) can be interpreted asymptotically
as,
\begin{equation}\label{Meps}
M(\varepsilon)\approx M_n=U_n/T_n, \mbox{ for large }n,
\end{equation}
 where
  $U_n$ is the total length of the
  $S_n$ in  $(4\varepsilon, 1-\varepsilon)$,
while   $T_n$ represents the total length of the
  $S_n$ in $X$.   For each finite $n$,
$U_n=\mu_u[S_n\cap(4\varepsilon,1-\varepsilon)]$ and $T_n=\mu_u[S_n\cap
X]$, can be
calculated numerically as the sum of the lengths of  subintervals
$A_{nk}$,
from Eq.~(\ref{amk}). Numerical evidence indicates that
the sequence of $M_n$
   converges in the limit $n\to \infty$, at least  for almost all
(w.r.t.
Lebesgue) $\varepsilon$.   However, for certain values of $\varepsilon$,
this is not true.
  For example, for $\varepsilon = \epsilon_1=1/6$ the sequence
$M_n$ does not converge, since
$M_{2n}=1/3$ and $M_{2n+1}=1/2$ for $n=0,1,2,\dots$.  We expect that
such
problem $\varepsilon$ are ``atypical" in that they have Lebesgue measure
zero,  due to the weak convergence of the measures
$\mu_n$ to $\mu_u$.
  We also observe that, in general, the convergence occurs faster
  for smaller values of $\varepsilon$.

 Similarly, the value of $M(\varepsilon)$ can be
expressed by a ratio of natural measures of two intervals:
\begin{equation}
 M(\varepsilon)=
 {\mu_s\bigl([4\varepsilon,1-\varepsilon]\bigr) \over
  \mu_s\bigl([2\varepsilon,1-\varepsilon]\bigr) },
\label{mm3}
\end{equation}
  since the unknown proportionality constant $c$, mentioned in Remark 1,
is
eliminated due to cancellation.

We now establish a function which gives the topological entropy
$h_T(\varepsilon)$ of the tent gap-map, in terms of $M(\varepsilon)$.
Therefore,
$h_T(\varepsilon)$ is easily approximated, which may be considered
surprising, given  that $h_T(\varepsilon)$ is a ``devil's staircase."

\smallskip
 {\bf Proposition 1: }
  {\it
The  topological entropy $h_T$  of the tent map
  with $\varepsilon$-gap equals,}
\begin{equation}
h_T(\varepsilon)=\ln [1+M(\varepsilon)].
\label{entr}
\end{equation}

\smallskip
Sketch of the proof \cite{MM}: Consider
the Ruelle-Bowen  transfer operator $L$,
 associated with the mixing dynamical system $f$, which acts on an a
continuous
density function $g:X\to [0,1]$
\begin{equation}
 L[ g(x)] := \sum_{y\in f^{-1}(x)} g(y),
\label{transfer}
\end{equation}
found in Ruelle's Perron-Frobenius Theorem, \cite{Bowen75}.
 When $\varepsilon$ is a binary rational, (computers can only store such
numbers), then there is a conjugacy of $f_\varepsilon$ to a subshift of
finite
type, which is exactly a topological Markov chain, for which Bowen's
version
of Ruelle's Perron-Frobenius Theorem is stated.  If $\varepsilon$ is not
a  binary rational,
the conclusion of the theorem still holds \cite{MM}.

The conclusion is therefore as follows.
The operator $L$ has the eigenfunction $\rho(x)$, $L\rho(x)=\lambda
\rho(x)$, and the adjoint
operator $L^*$ has the eigenmeasure $\nu$, $L^*\nu=\lambda \nu$ which
may be
normalized so that $\nu(X)=1$; both $\rho(x)$, and $\nu$ correspond to
the same
largest eigenvalue $\lambda$, and the maximal entropy measure $\mu_*$ is
uniquely absolutely
continuous with respect to $\nu$ by the density function $\rho(x)$.
Thus
$d\mu_*(x)=\rho(x)d\nu(x)$.
The topological entropy of $f$ is equal to the logarithm of this largest
eigenvalue, $\lambda$, of $L$ and $L^*$ \cite{Bowen75}.

For the model Eq.~(\ref{ten}), there is constant slope $|f'|=2$,
and therefore, for arbitrary $n$, any normalized uniform measure
$\mu_n(X)=1$ from Eq.~(\ref{uniformn}) iterates,
\begin{equation}\label{L*}
L^*[\mu_n]=m_n \mu_{n+1},
\end{equation}
 where $m_n\equiv 1+M_n$, and $\mu_{n+1}$ is also uniform.
Therefore, in the weak limit
$n\to \infty$, we obtain $L^*[\mu_u]=\lambda \mu_u$.  Thus, the measure
$\nu$ is equal to
the uniform measure $\mu_u$, and both are absolutely continuous to the
maximal
entropy measure $\mu_*$.

The measure of the
 entire space $X$ with respect to $L^*(\nu)$ is given by
\begin{equation}
 L^* \nu (X) = \nu(E_1) +2 \nu(E_2),
\label{measx1}
\end{equation}
since, by definition, for any $x\in E_2$ there exist two preimages
$f^{-1}(x)\in S$,
and  for any $x\in E_1$ there exist only one such preimage.
Making use of the eigenequation,
 $L^* \nu =\lambda \nu$, and the
normalization condition $\nu(E_1)+\nu(E_2)=1$ we arrive at
\begin{equation}
  \lambda  = 1 + \nu(E_2),
\label{measx2}
\end{equation}
which proves the formula (\ref{entr}), making use of our special case
result
that  $\nu=\mu_u$. $\Box$

\section{ Direct approximation of the topological entropy function}

In this section, we give numerical and graphical
 evidence as to the accuracy
of Proposition 1.

\smallskip
{\bf Example 1:}
For $\varepsilon=1/7$ the
continued fraction expansion of (several) numbers $M_n$ consists of
ones only, and the sequence $M_n$ converges to the golden number:
 $M(\varepsilon=1/7)=(\sqrt{5}-1)/2=\gamma \approx 0.618034$, so that
 $h_T (\varepsilon=1/7)= \ln[(\sqrt{5}+1)/2]\approx 0.481212$.

\smallskip
{\bf Example 2:}
We now consider a sequence of closed form approximates  to the
   devils staircase entropy function $h_T(\varepsilon)$, based on
Eqs.~(\ref{Meps})-(\ref{entr}).
As the zero order approximation of $M$, we take the ratio
between Lebesgue
 measures $M_0=(1-\varepsilon-4\varepsilon)/
(1-\varepsilon-2\varepsilon)$. Substituting
this into (\ref{entr}) gives for $\varepsilon \in [0, 1/5]$
an analytical approximation of the entropy
\begin{equation}
h_0(\varepsilon) = \ln \Bigl[{2-8\varepsilon \over 1-3\varepsilon}
\Bigr]   ~ ~ {\rm for} ~ ~ \varepsilon \in [0,1/5]
\label{a0}
\end{equation}
represented in Fig.~2 by a thin solid line. This formula gives
the asymptotic behavior $h_T(\varepsilon)\approx \ln2-\varepsilon$,
valid for small $\varepsilon$.

\smallskip

{\bf Example 3:}
Calculating higher order ratios
$M_n$ yields better approximations for the topological
entropy function. For example, $M_1$ gives
\begin{equation}
h_1(\varepsilon)  =  \left\{ \begin{array}{lll}
  \ln \bigl[ (2-10\varepsilon)/ (1-4\varepsilon) \bigr]
   &~\hbox{for}~&  \varepsilon   \in [0, 1/9]  \nonumber \\
  \ln \bigl[ (3-11\varepsilon) /(2-8\varepsilon)  \bigr]
   &~\hbox{for}~&  \varepsilon   \in [1/9, 1/7]   \\
  \ln \bigl[ (2-9\varepsilon) / ( 1-4\varepsilon)  \bigr]
   &~\hbox{for}~&  \varepsilon   \in [1/7, 1/5] \nonumber \\
\end{array}
\right.
\label{a1}
\end{equation}

while the next approximate, based on $M_2$,  reads

\begin{equation}
 h_2(\varepsilon)  =  \left\{ \begin{array}{lll}
  \ln \bigl[ (2-12\varepsilon)/ ( 1-5\varepsilon)  \bigr]
   &~\hbox{for}~&  \varepsilon   \in [0, 1/17]  \nonumber \\
  \ln \bigl[ (7-31\varepsilon )/ ( 4-20\varepsilon) \bigr]
   &~\hbox{for}~&  \varepsilon   \in [1/17,1/15]  \nonumber \\
  \ln \bigl[ (4-23\varepsilon) /  ( 2-10\varepsilon)  \bigr]
   &~\hbox{for}~&  \varepsilon  \in [1/15,1/9]  \\
  \ln \bigl[ (5-19\varepsilon) / (3-11\varepsilon)  \bigr]
   &~\hbox{for}~&  \varepsilon  \in [1/9,1/7] \nonumber \\
  \ln \bigl[ (4-20\varepsilon) / ( 2-9\varepsilon)  \bigr]
   &~\hbox{for}~&  \varepsilon  \in [1/7, 2/11]. \nonumber \\
\end{array}
\right.
\label{a2}
\end{equation}
The $h_1(\varepsilon)$ and $h_2(\varepsilon)$ approximates can
also be found in Fig.~2, and we see that $h_2(\varepsilon)$
 is already pretty good.
  The solid thick line  represents
$h_{10}(\varepsilon)$, obtained with the uniform measure supported
on the
set $S_{10}$, by numerically computing the ratio $M_{10}$, according to
Eq.~(\ref{Meps}). Crosses denote exact results, computed by roots of
polynomials, as described in Appendix A, for
values of $\varepsilon$ corresponding to periodic orbits of length $L
\le 6$. When $\varepsilon <1/7$, these results
coincide to within $10^{-5}$. On the other hand some discrepancies are
visible for larger gaps, for which convergence of the
sequence $M_n$ is slower.

\smallskip
{\bf Conjecture:} Iterating this procedure produces continuous,
piecewise differentiable functions $h_n(\varepsilon)$
 which, in the limit $n\to \infty$,  converge to
the topological entropy versus noise gap function
$h_T(\varepsilon)$. Convergence is in the $L^1$--norm, and
$\int_0^1 |h_T(\varepsilon) - h_n(\varepsilon)| d\varepsilon
\to 0$ as $n\to \infty$.

\section{Structure of the entropy devil's staircase}

In fact, Eqs.~(\ref{Meps})-(\ref{entr}) of Proposition 1, together with
the
closed form representation of $A_{mk}$, found in Eq.~(\ref{amk}) can be
used
to investigate the structure of the devil's staircase topological
entropy
function, $h_T(\varepsilon)$, which we do in this section.

Our primary remark of this section follows from the observation that the
function
    $M(\varepsilon)$ is constant  for $\varepsilon \in \{
\varepsilon: 4\varepsilon \in  A_{mk}\}$ (for any $m$ and $k$),
since varying the parameter $\varepsilon$ both integration
borders sweep the empty region of $X$ and the measure
 $\mu_u(\varepsilon)$ does not change in each of these intervals.  This
is
equivalent to an alternative topological description of the same
phenomenon:
when word bins overlap, no
 change takes place to the topological entropy of
the corresponding symbol dynamics,
which must be a subshift of finite-type
\cite{bollt-lai2}. In other words, varying the parameter
$\varepsilon$ in these regions does not influence the set of periodic
orbits, and hence the
 topological polynomials \cite{MT77},  or dynamical
zeta functions \cite{Ru78}, are unchanged and therefore lead to
the constant topological entropy.

The condition that the integration
``sweeps a gap" when $\varepsilon \in \{
\varepsilon: 4\varepsilon \in  A_{mk}\}$,
defined in (\ref{amk}), gives
\begin{equation}\label{Bmk}
   M(\varepsilon)={\rm const ~ ~ ~ for} ~~
    \varepsilon \in B_{m,k}:= \Bigl[ {2k-1 \over    2^{m+3}+1},
        {2k-1 \over    2^{m+3}-1}\Bigr],
\label{cons}
\end{equation}
  for each $m$-pre-iteration of the gap
  (word-length $m$) ($m=0,1,2,\dots$
and each $m$-bit word, $k=1,2,\dots,2^m$.

In particular, we have monotonicity of
 the $M(\varepsilon)$ function, and so
it must follow that
 \begin{equation}
M(\varepsilon_1) \le M(\varepsilon_2) \mbox{ for } \varepsilon_1 >
\varepsilon_2.
\end{equation}
It therefore immediately follows that topological entropy
$ h_T(\varepsilon)$ is a
non--increasing function of $\varepsilon$, as was proven in
\cite{BMT91}.

\smallskip
{\bf Example 4:}
The gap $A_{01}$ generates the
 main plateau of the ``devils staircase"
$B_{0,1}=[1/9,1/7]$, for which the topological entropy
is equal to $\ln[(\sqrt{5}+1)/2]$ - see Example 1 above.

\smallskip

\smallskip
{\bf Remark:} Inside the main plateau $B_{0,1}=(1/9,1/7)$
all three of the approximate functions, $h_0(\varepsilon)$,
$h_1(\varepsilon)$ and $h_2(\varepsilon)$, cross at
$\varepsilon=(6-\sqrt{5})/31\approx
0.121417$, for which the sequence $M_n$ is constant and equal to
the golden mean $\gamma$, corresponding to the exact value of the
entropy $h_T=$ln$(1+\gamma)$.  See Fig.~2.

\section{Kneading sequences  and the entropy plateaus}

The kneading  theory of Milnor and Thurston \cite{MT77} allows us
to compute the topological entropy of a unimodal map from the orbit of
its critical
point, using the so--called kneading determinant (see e.g.
\cite{MeloStrein}).
  This technique may also be applied in our analysis of the tent map
with a symmetric gap, for which the critical orbit originates in
  one of (either) two ends of the gap: $[(1-\varepsilon)/2,
(1+\varepsilon)/2]$.
It enables us to express
the topological entropy for any
flat steps of the staircase, $\varepsilon \in B_{mk}$, as
\begin{equation}\label{h_Temk}
h_T(\varepsilon_{mk})=\ln(\lambda_{mk}),
\label{poln}
\end{equation}
 where $\lambda_{mk}$ is the {\sl largest} (real) root of the polynomial
$P_a(z)$ of order  $m+2$,
\footnote{added line}
which is the characteristic polynomial of the
Stefan transition matrix \cite{DGP78}.
As shown in  Appendix A, all of the coefficients
$ [c_{m+2}, c_{m+1},\dots,c_1,c_0 ] $ are
equal to either $+1$ or $-1$. These coefficients are uniquely determined
by the kneading
sequence, representing the symbolic itinerary of a periodic critical
orbit.  It follows, from Eq.~(\ref{h_Temk}), that the
values  of the corresponding plateau $B_{mk}$ in the
space of $\varepsilon$, are also uniquely determined by these same
coefficients.

We remark  that for the polynomials $P_a(z)$ are
closely related to the kneading determinant
 $P_a(z)$ of Milnor and Thurston. For any plateau
$B_{mk}$, associated with an orbit of
the length $L=m+3$, the kneading invariant is proportional
to a finite polynomial $P_b(1/z)$
 and the topological  entropy is given as the logarithm of the
{\sl smallest} root of $P_b(1/z)$ \cite{MT77,Gu80}. Kneading
determinants are considered a standard tool
 to compute the topological entropy of one-dimensional maps
\cite{CCE83,HK85,CCP95a}, but for our purposes, we find the related
polynomials $P_a(z)$ are more convenient.

To highlight our understanding of the relationship between the function
$h_T(\varepsilon)$,
displayed in Fig.2, relative to calculations based on
Eq.~(\ref{h_Temk}), we have constructed Table 1, by collecting the
kneading sequences, corresponding polynomials, their
largest roots, and
the values of the topological entropies for plateaus corresponding
to periodic orbits.
Each plateau occurs for $\varepsilon \in [\varepsilon_- ,
\varepsilon_+]$ given by Eq. (\ref{cons}).
The orbits of length $L=3 \mbox{ through }7$ (and some of the length
$L=8$) are ordered
according to the decreasing entropy, which corresponds to increasing
width of the gap $\varepsilon$. Any periodic orbit may represent a
kneading sequence, but not vice-versa; some kneading sequences (and
hence the corresponding polynomials) are not admissible
for the tent  map (see eg. \cite{De89,HB89}) as they do not
correspond to any
periodic orbits in the system, and therefore, do not affect
the dependence $h_T(\varepsilon)$.

\smallskip

{\bf Example 5:} Consider the main plateau $B_{0,1}=(1/9,1/7)$. The
critical orbit is has the length $L=m+3=3$,
representing the kneading sequence {\bf CRL}, corresponding to the
sequence of coefficients $[+ - -]$, which denotes the polynomial
$P_a(z)=z^2-z-1$.
The symbols {\bf C, R, L}, are used to mark, whether  each iterate is at
the critical point (Center), right of it (Right) or left of it (Left).
Largest root of the (this) equation $P_a(z)=0$ equals
$\lambda_{01}=1+\gamma$ where $\gamma$ denotes the golden mean
$(\sqrt{5}-1)/2$.  The one follows the same
result for topological entropy discussed in Proposition 1.

\smallskip

We characterize a peculiarity in the structure of the devil's staircase,
visible in Fig.1,  by the following

\smallskip

 {\bf Proposition 2:}
  {\it
Any entropy plateau corresponding to an orbit of length $L$ is
accompanied on the left (smaller gap width) by an infinite number
of adjacent plateaus with the same entropy, which are caused by orbits
of period $2L$, $4L$, $8L,\dots$. }
\smallskip

To prove this proposition, consider the sequence of signs
$[c_{L-1},c_{L-2}\dots,c_1,c_0]$, (defining a polynomial corresponding
to an orbit of length $L$, as mentioned above). Also consider the
following operations acting on the sequences of
signs $c_i=\pm 1$ of length $L$, which double their length:
\begin{equation}
W_1[c_{L-1},c_{L-2}\dots,c_1,c_0] =
   [c_{L-1}, c_{L-2}, \dots,c_1,c_0,-c_{L-1},
    -c_{L-2}\dots,-c_1,-c_0],
\label{W1}
\end{equation}
\begin{equation}
W_2[c_{L-1},c_{L-2}\dots,c_1,c_0] =
   [c_{L-1},c_{L-2}\dots,c_1,c_0,c_{L-1},
    c_{L-2}\dots,c_1,c_0],
\label{W2}
\end{equation}
In a straight forward way, these transformations, discussed in
\cite{MT77},
 may be mapped into
the space of polynomials, of order $L-1$, with all coefficients equal to
$\pm 1$
\begin{equation}
W_1\bigl(P_a(z)\big) =(z^{L}-1) P_a(z); ~~~~~
W_2\bigl(P_a(z)\big) =(z^{L}+1) P_a(z).
\label{W12P}
\end{equation}
Since the roots of the factor $(z^L \pm -1)$ are situated
on the unit circle, the largest real roots of the polynomial
$\alpha=P_a(z)$, and its two images,
$\alpha'=W_1\bigl(P_a(z)\bigr)$ and
$\alpha''=W_2\bigl(P_a(z)\bigr)$, are the same.
Therefore, if there exist admissible periodic orbits corresponding to
the
``descendent'' polynomials $\alpha'$ and
$\alpha''$, then they form a plateau of the same height as their
``ancestor"
polynomial $\alpha$.

We find, in fact, that there are such plateaus corresponding to the
descendants $\alpha'$, but not for the descendants $\alpha''$.
We establish, in a somewhat indirect way, the existence of the $\alpha'$
plateaus by
considering their location
(terminology as used in \cite{De89}). Let us rewrite,
in a simplified form, the position of the plateau induced by the
polynomial
$\alpha$ according to Eq. (\ref{cons}):
$B_{L,j}=[j/(2^L+1), j/(2^L-1)]$, where $L=m+3$ and $j=2k-1$.
Using the fact that the sequence of coefficients $c_i$ represent the
integer $k-1$ in the binary code (but not the itinerary code: see
Appendix A), we arrive
at the conclusion that the polynomial $\alpha'$ is associated with the
plateau
$B'=B_{2L,j'}$, where $j'=j(2^L-1)$. Therefore, this descendent
plateau
$B'=[j(2^L-1)/(2^{2L}+1), j(2^L-1)/((2^L-1)(2^L+1))]$ touches, from
the left, the ancestor plateau  $B_{L,j}=[j/(2^L+1), j/(2^L-1)]$ and
thus influences the devil staircase.
This reasoning is valid for any admissible periodic orbit of
length $L$. Since the descendent plateau, corresponding to the orbit of
length $2^n L$, has descendants related to the orbit of the length
$2^{n+1} L$,
there exists an infinite sequence of plateaus (related to the orbits
of length $2^nL$, $n=1,2,\dots$) and
Proposition 2 is justified. Furthermore, the length of these adjacent
plateaus,
determined by the denominators $2^n L$, decreases exponentially with
$n$, and their total sum gives the total width of a plateau.

The ``would-be" $\alpha''$ plateaus may be formally constructed, but we
find that they are entirely included within the boundaries of the
ancestor plateaus $\alpha$.
In the analogous construction to that of the previous paragraph, we find
that the polynomial $\alpha''$ would be associated with plateau
$B''=B_{2L,j''}$, where $j''=j(2^L+1)$. Calculating the location,
$B''=[j(2^L+1)/(2^{2L}+1), j/(2^L+1))]$, we see that its right edge
coincides with the right edge of the
plateau  $B_{L,j}$. I.e., $B''\subset B_{L,j}$.  Consequently,
descendants $\alpha''$ are
entirely shadowed by the longer ancestor plateaus and do not
affect the devil staircase. This reflects the fact that the sequences
$\alpha''$ do not correspond to any admissible periodic orbits
(see eg. \cite{HB89}, pp. 136-139).

\smallskip

{\bf Example 6:} The golden plateau $B_{0,1}$, with $\varepsilon\in
[1/9,1/7]$ found by Eq~(\ref{Bmk}), is represented by the
polynomial $\alpha=[+ - -]$, according to
Eqs.~(\ref{coeffs0})-(\ref{coeffs}), using $m=0$, $k=1$ and $L=m+3=3$.
The two descendant polynomials are $W_1(\alpha)=\alpha'=[+ - - - + +]$
and
$W_2(\alpha)=\alpha''=[+ - - + - -]$. The former represents the
orbit {\bf CRLLRL}
leading to the plateau for $\varepsilon \in B_{3,4}=[7/65, 7/63]$ (again
by Eqs.~(\ref{Bmk}) and (\ref{coeffs0})-(\ref{coeffs}),
which forms an extension of the ``golden" plateau $[1/9,1/7]$ of the
same
entropy ln$(1+\gamma)$. The latter corresponds to the
non-admissible orbit  {\bf CRLRRL} \cite{CCE83}
and the hidden plateau $B_{3,5}=[9/65,9/63]$ which is a subset of
the golden  plateau.
 Existing left extensions of shorter orbits plateaus are
marked by ``E" in the Table 1, and for pedagogical purposes, we also
include the non-existing hidden plateaus (stemming from polynomials
$\alpha''$)
marked by the letter ``N".

\smallskip
In order to analyze the case of wide $\varepsilon$--gaps,
characterized by a decreasing entropy, it is helpful to
consider two other operations doubling the sequences of signs
\begin{equation}
U_1[c_{L-1},c_{L-2}\dots,c_1,c_0] =
   [c_{L-1},-c_{L-1}, c_{L-2},-c_{L-2}, \dots,c_1,-c_1,c_0,-c_0],
\label{U1}
\end{equation}
\begin{equation}
U_2[c_{L-1},c_{L-2}\dots,c_1,c_0] =
   [c_{L-1},c_{L-1}, c_{L-2},c_{L-2}, \dots,c_1,c_1,c_0,c_0],
\label{U2}
\end{equation}
and the corresponding transformations in the space of polynomials
\begin{equation}
U_1\bigl(P_a(z)\big) =(z-1) P_a(z^2); ~~~~~
U_2\bigl(P_a(z)\big) =(z+1) P_a(z^2).
\label{U12P}
\end{equation}
Let $\lambda_{\alpha}$ denotes the largest root of the polynomial
$\alpha$ of order $L-1$. It is easy to see that the largest roots of the
descendent polynomials $\alpha_3=U_1(\alpha)$ and $\alpha_4=U_2(\alpha)$
of order $2L-1$ are equal to $\sqrt{\lambda_{\alpha}}$,
so the corresponding entropies are halved. The sequences $\alpha_4$ do
not correspond to any of the admissible periodic orbits \cite{HB89},
and the operator $U_2$ is mentioned here for completeness only.
On the other hand, a {\sl renormalization} operator $U_1$,
generating admissible periodic orbits $\alpha_3$,
is often discussed in the literature \cite{CCE83,HB89,CCP95a}. In a
natural way
this operator can be generalized to act in the space of infinitely long
sequences. The corresponding operation of the kneading sequences,
 which doubles the length of the periodic orbit,
 is a special case of the Derrida-Gervois-Pomeau $*$ composition
\cite{DGP78}.

\smallskip

{\bf Remark:} We are now in
 a position to bound the critical last gap value
$\epsilon_*$, for which any larger $\varepsilon$-gap has no topological
entropy, since all of the gaps have overlapped. This remark is
summarized by Fig.~3.  The tent map with  no
gap $(\varepsilon=\epsilon_0=0)$ is characterized by the kneading
sequence
{\bf Q=CR(L)$^{\infty}$}, polynomial $\alpha_0=[+ - - - \cdots]$ and the
entropy ln$2$. The kneading sequence {\bf R*Q=CRL(R)$^{\infty}$}
is represented by the polynomial
$U_1(\alpha_0)=[+ - - + (- +)^{\infty}]$. Consequently, the
entropy ln$2/2$ is achieved for $\varepsilon = \epsilon_1=1/6$.
In this way we construct a family of
kneading sequences {\bf R$^{n}$*Q} and the polynomials
$U_1^n(\alpha_0)$, which allow us to find the sequence
of numbers $\epsilon_n$ such that
$h_T(\epsilon_n)=2^{-n}\ln 2$. In particular
$\epsilon_2=7/40=0.175$, while already
the next value $\epsilon_3 \approx 0.175092$
provides a good approximation of the Feigenbaum critical point
$\epsilon_*=\lim_{n\to\infty} \epsilon_n$.

The same value can be approached from above by considering
wider gaps $\varepsilon > \epsilon_*$, corresponding
to periodic orbits of the length $L=2^l$, which lead to the
zero entropy. A gap of the width $\varepsilon=e_1=1/5$
leads to the orbit {\bf S=CR} and a trivial polynomial $\beta_1=[+ -]$
with the root $\lambda=1$. For this polynomial both operations
$W_1$ and $U_1$ produce the same result $\beta_2=[+ - - +]$
(since half of zero entropy is equal to zero). The corresponding orbit
{\bf CRLR} appears at $\varepsilon=e_2=3/17 \approx 0.17647$.
Subsequent processes of period doubling occur at the gaps
$\varepsilon=e_n$
corresponding to the polynomials $\beta_n=U_1^n(\beta_1)$. For example
$\beta_3=[+ - - + - + + -]$ gives $e_3=45/257\approx 0.175097$,
 where $L=8$, $m=L-3=5$,
$k=2j+1=45$,  $j=22$, and $2^L+1=257$ are all consistent, again by
Eqs.~(\ref{Bmk}) and (\ref{coeffs0})-(\ref{coeffs}). In general
\begin{equation}
e_n =  {1 \over 2^{2^n}+1 } \prod_{k=0}^{n-1} \bigl( 2^{2^k} -1 \bigr),
\label{el}
\end{equation}
and each zero entropy plateau $\varepsilon \in [e_{n+1},e_n]$ forms an
extension of the plateau $[e_n, e_{n-1}]$.

The sequence $e_n$ converges quadratically ($ (e_{n+1}-e_n) \approx
(e_n-e_{n-1}) ^2 $),
 in contrast
with the geometric convergence of the
well known Feigenbaum sequence ($(e_{n+1}-e_n)/(e_n-e_{n-1}) = \delta$,
 asymptotically as the Feigenbaum delta constant), which describes the
period doubling in the logistic
map \cite{Ott94}.  The
first $15$ decimal digits of $e_5$ and $e_6$ are the same
and provide an excellent approximation of the
Feigenbaum point,
\begin{equation}
 e_*=\epsilon_*\approx   0.17509193271978.
\end{equation}
This can be considered as a sharp estimate for the ``maximal"
topologically nondegenerate gap.
A sketch of the behavior
of the function $h_T(\varepsilon)$ in the vicinity of the critical
point $e_*$ is shown in Fig.~3.

 Previous attempts to calculate the
critical last gap $\varepsilon$, by ``brute-force" direct computation of
the
topological entropy (by counting symbol sequences) on the invariant set,
break-down due to the exponentially increasing difficulty (see
conjecture
associated with Eq.~(\ref{rate}) of locating the ever thinning
invariant set
(by PIM triple method) \cite{bollt-lai2}.  For the first time, we can
now
draw the curve all the way to zero entropy, as shown in Fig.~3.

\smallskip

{\bf Remark:}
 The analyzed structure of the dependence of the
   topological entropy is typical for all unimodal maps with a gap.
   While location  of the entropy plateaus
  depends on the map and on the position of the gap,
  the heights of the plateaus are universal. The same sequences
  of entropy plateaus occurs for some 1D maps {\sl without} the gap.
  For example, the periodic windows for the logistic map are
characterized by zero
(or negative) KS-entropy and a constant topological entropy. Its value
is
 determined just by the kneading sequence of the periodic orbit
 and therefore, can be expressed by roots
 of the polynomials discussed in this work.
 A figure of parametric dependence of the topological
entropy for quadratic map (related with logistic map)
 is already sketched in the preprint of Milnor and
Thurston \cite{MT77}, and a more precise picture of such dependence is
 presented in the review of Eckmann and Ruelle \cite{ER85}
and in the paper by G{\'o}ra and Boyarsky \cite{GB91}.

\section{Topological Entropy is the ``Average" Pre-Image Count}

In this section, we pursue a result, suggested by part of the proof of
Proposition 1.  For a more general map than the (constant slope) tent
map, we
do not have the coincidence of uniform measure $\mu_u$ and the Parry
measure
$\mu_*$, and therefore we cannot  conclude Eq.~(\ref{L*}) with a uniform
$m_n$.  Nonetheless, we can make the following general Proposition,

\smallskip
{\bf Proposition 3:}
 {\it   The topological entropy of a 1D mixing system $f: X\to X$, for a
piece-wise
monotone function $f$, which is continuous on the $N$ branches,  is
equal to
\begin{equation}
 h_T  = \ln \int_X  P(x) d \nu(x),
\label{h00}
\end{equation}
where $P(x):X\rightarrow \{0,1,2,...,N\}$ represents the number of
preimages
of $f$ at the point $x$
(restricted to the support of $\nu$)
and the average is taken with respect to the $L^*$ eigenmeasure measure
$\nu$, which is absolutely continuous to the maximal entropy measure
$\mu_*$ by the $L$ eigendensity $\rho(x)$.
}

\smallskip Note that one cannot generally expect  P(x) to be surjective,
onto the set
$\{0,1,2,...,N\}$.

\smallskip
Proof: The proof is very similar to the second half of the proof of
Proposition
1, which is all that survives the weaker condition, that we allow maps
with
arbitrary slopes.  As before, we split $X$ into $X=\cup_{j=0}^N E_j$,
where
$E_j=\{x:x\in X \mbox {, and }f^{-1}(x) \mbox{ has }j \mbox{ branches
}\}$.
The adjoint eigenstate equation of the Bowen transfer operator $L^*$,
measuring the whole space $X$, is
\begin{equation}\label{preimages}
L^*\nu(X)=\sum_{j=0}^N j \nu(E_j)=\int_X P(x) d\nu(x)\equiv
<P(x)>_{\nu}=
<\frac{P(x)}{\rho(x)}>_{\mu_*},
\end{equation}
where, as before, the eigenmeasure of the operator $L^*$, $\nu$, is
known to be
uniquely absolutely continuous to Parry's maximal entropy measure
$\mu_{*}$, by
$\rho(x)$, which is the eigenfunction of the adjoint eigenequation
$L\rho=\lambda
\rho$.  Therefore, the eigenvalue of this equation
is $\lambda=<P(x)>_{\nu}$,
and the topological entropy is,
$h_T=\ln(<P(x)>_{\nu})=ln<\frac{P(x)}{\rho(x)}>_{\mu_*}$. $\Box$

\smallskip
{\bf Example 7:} Take any unimodal map with a.e.~2 preimages, such as
the
logistic
 map, $x'=4x(1-x)$ which is well known \cite{Ott94} to
have topological entropy $h_T=ln(2)$, when the parameter value $a=4$.
This result is particularly easy to derive by Proposition 3, for which
we may check that, $h_T=\ln(<P(x)>_{\nu})=ln(\int_{[0,1]}2
d\nu)=ln(2\int_{[0,1]} d\nu)=ln 2$, as expected.  The main simplifying
feature of the calculation is that only the number of branches,
weighted by the normalized measure $\nu$, was important, and thus the
calculation is quite general (and hence identical for, say, the
two-onto-one tent map, the two-onto-one cusp map, etc.).

\smallskip
We note that our formula Eq.~(\ref{preimages}) is reminiscent to a
similar
formula, for almost all $x$, \newline$h(f)=\lim_{n\rightarrow}
\frac{1}{n} log
\#f^{-n}(x)$,   used in Lopes and Withers, \cite{LopesWithers}.  The
link
follows by considering their formula as an average of the number of
pre-images, of the initial condition, in which some branches are
presumably
dense in $X$, which is just the same as using the measure of maximal
entropy.

\section{Fractal Dimension and Escape Rate}

So far, we have only considered the topological entropy.  The so-called
flat-spots of the topological entropy function, caused by overlapping
symbol
bins, which causes the integral
of Eq.~(\ref{mm}) for $(M(\varepsilon)$ to ``sweeps a gap" when
$\varepsilon
\in \{
\varepsilon: 4\varepsilon \in  A_{mk}\}$ defined in (\ref{amk}), also
has
consequences to the spectrum of  pointwise spectrum of Renyi-Dimensions
$D_q$ \cite{Renyi2} (see e.g.~\cite{Ott94} pp 79, or pp 306 for review).
We now discuss these implications in this section.

We conjecture that the Hausdorff dimension of the support $S$ (and the
measure $\mu_u$) coincide with all of the
generalized Renyi (multifractal) dimensions $D_q$ \cite{Ott94},
and hence we write
\begin{equation}
  D_0=D_q=   { \ln ( 1+M) \over \ln 2 } .
\label{dim1}
\end{equation}
In \cite{BR87}, we find the relationship $D_1=h_{KS}/\ln 2$, directly
linking
the {\it information dimension} proportionally to the KS entropy, and
this
corresponds
to the Kaplan-Yorke conjecture \cite{Ott94}, formulated in a different
setting. The dimension is thus proportional to entropy and displays the
same devil staircase like dependence on the parameter $\varepsilon$.

\smallskip
Next we consider the nature of this map as a dynamical system on the
unit
interval, whose invariant set is an unstable chaotic saddle.
Therefore, the initial conditions which are not on this invariant set,
 escape to infinity.
In the analogy to \cite{BR87,JacobsOttHunt}, we conjecture that
iterating
$f_\varepsilon$, on an initially uniform measure,
 causes the mass of points to
decay exponentially with the number of iterations, according to
exp$(-Rn)$.
 From such an exponential decay
model follows the exponent $R$,
\begin{equation}
 R=\ln 2 - h_{KS}= \ln \Bigl[ {2\over 1+M} \Bigr].
\label{rate}
\end{equation}
Furthermore, this escape rate $R$ describes the
exponential convergence of the series $M_n$, in Eq.~(\ref{Meps}).
 Note that the number $M(\varepsilon)$,
 in Eq.~(\ref{mm}), defines the limit of this sequence, as well
its convergence rate.

There is a striking similarity between the topological entropy
devil's staircase function of $f_\varepsilon$, and the similar
devil's
staircase topological entropy of the logistic map $f_r=rx(1-x)$ on the
parameter $r$ \cite{CCP95a,CCP95b}.  This follows immediately from
the fact
that in both models, we are monotonously nonincreasing the
kneading sequence, with the parameters $\epsilon$ and $r$ respectively.
However, in the case of the tent gap-map $f_\varepsilon$, the set $G$ of
$\varepsilon$ values which are  {\sl not} contained in the flat
steps $B_{mk}$  is of zero Lebesgue  measure \cite{BMT91}
and has a fractal structure \cite{JacobsOttHunt}.
 In contrast, the set
of $r$ values which lead to chaotic motion (not
contained  in the ``periodic windows" of a constant topological entropy)
has a positive Lebesgue measure \cite{Ja78}.

It is natural to investigate the
 homogeneity and the local pointwise dimension $D_l$ of
the set $G\subset[0,\varepsilon_*]$ of $\varepsilon$ values {\sl not}
contained in the flat steps $B_{mk}$.
 Consider a fixed value of $\varepsilon$ and a set $S$ of
dimension $D_{\varepsilon}$ which supports the
invariant measure $\mu_*(\varepsilon)$ of the system. We perturb
the size of the gap, $\varepsilon'=\varepsilon+\delta$, and
we find that in the limit $\delta\to 0$ the measure
$\mu_*(\varepsilon')$
converges weakly to $\mu_*(\varepsilon)$.
As we have already discussed that Eq.(\ref{entr}) of Proposition 1
implies that
entropy changes only if
the integral (\ref{mm}) changes, which occurs as
the integration borders sweep across the fractal set $S$, but not when
we sweep the gaps $\bar{S}$.

Therefore we conjecture that the set $G$ is nonhomogeneous
and its local point
 dimension $D_l$ depends on the size of the gap according to
\cite{JacobsOttHunt},
\begin{equation}
D_l(G(\varepsilon))= D(\varepsilon)= {\ln \bigr(1+M(\varepsilon)\bigl)
\over \ln 2 }.
\label{dimg}
\end{equation}

\smallskip
{\bf Remark.} The intervals
$\Delta_n:=[\epsilon_{n},\epsilon_{n+1}]$ are {\sl
similar} in the sense that
any plateau in the interval $\Delta_0$, associated with the polynomial
$\alpha$, has a corresponding plateau in each of the
intervals $\Delta_n$ and these descendent plateaus are represented by
polynomials
$U_1^n(\alpha)$.

\smallskip

{\bf Example 8.} An orbit {\bf CRLRRR} is associated with the polynomial
$[+ - - + - +]= U_1([+ - -])$, so the corresponding plateau is localized
in $\Delta_2$ and its entropy is equal to $[\ln (1+\gamma)]/2\approx
0.240606$. Another descendent plateau, determined by $U_1([+ - - -])$,
and corresponding to the orbit of the length $8$, is marked in
Table 1 by the letter ``D".

\smallskip

Despite the similarity emphasized above, the entropy devil's
staircase is not self--similar
in the intervals $\Delta_n$. It is not possible to linearly
rescale the interval $\Delta_n$ by a constant factor
to get the dependence $h_T(\varepsilon)$ in the next interval
$\Delta_{n+1}$. This corresponds to the fact
that the set $G$ is not homogeneous and its local dimension
varies with $\varepsilon$.

\smallskip
{\bf Remark.} The number $I_L$ of plateaus generated by periodic
orbits
of the length $L$ in the first interval $\Delta_0$  are listed in
Table
2. We do not count those orbits, which produce plateaus embedded
in longer plateaus generated by shorter orbits.
The column $T_L$ represents the total width of all plateaus generated by
all orbits of the length $L$, while the last column  $W_L$ represents
the total volume of the parameter space in $\Delta_0$ not contained
in the sum of the plateaus $T_L$.
A naive exponential fit gives  $W_L=a+be^{-cL}$
with a positive $a\sim 0.03$, but if $G$ is indeed a fractal contained
in the interval, then
this approximation can not be true since $a$ should be zero.
Comparison with a similar table obtained
for logistic map \cite{CCP95b} shows that the number of flat steps
in the
entropy dependence on the parameter, which correspond
to periodic orbits of a fixed length, are almost the
same. On the other hand, the relative Lebesgue measure of
the plateaus in the parameter space
is much smaller for the logistic map.

\section{Pragmatic Conclusions}

\smallskip
We have performed a detailed measure theoretic based analysis of the
devil's staircase topological entropy function of the gap-tent map whose
invariant set is an unstable chaotic saddle invariant set of the tent
map.  The point was to further analyze the trade-off between channel
capacity and noise-resistance in designing dynamical systems to pursue
the idea of
communications with chaos.  One may reduce the
effects of an external noise by
introducing a gap into the system (i.e. by not using parts of the
phase space  close to the partition lines).
We explicitly demonstrate that
some levels of noise are better than others for this purpose.
 For the simple tent-gap map model system ($2x$ tent map) the noise gap
$\varepsilon=1/7$
 provides the same maximal information transmission rate
(topological entropy) as the gap $\varepsilon=1/9$ and offers $128\%$
larger immunity against  noise.
In general, for this system the gaps
   $\varepsilon = (2k-1)/ (2^{m+3}-1)$
(at the right edges of the plateaus $B_{mk}$),
 are more useful
than when $\varepsilon = (2k-1)/ (2^{m+3}+1)$
(at the left edges of the plateaus),
with fixed natural numbers $m$ and $k$.
Our analysis can also be applied to investigate
the effects of noise in measurements
performed by electronic devices, in which
the result of measurement is determined by a
symbolic sequence describing a chaotic trajectory
\cite{Ke95}.

We would like to mention, in Appendix B, a brief description for our
future research, by which the measured statistical properties of
deterministic dynamical systems are linked to an appropriately chosen
stochastic system by the so-called iterated function systems theory.

\section*{Acknowledgments}

We are indebted to M.~Misiurewicz for a particularly
helpful hint leading to the proof of Proposition 1.
We also thank Y.~Jacobs, B.~Hunt, D.~Lathrop, M.~Ogorza{\l}ek,
E.~Ott, T.~Kapitaniak,
W.~S{\l}omczy{\'n}ski, E.~Withers, J.A.~Yorke and G.-C.~Yuan for
helpful discussions.
K. {\.Z}. acknowledges the Fulbright Fellowship and a
support by Polish KBN grant no. P03B~060~13.
E.~B. is supported  by the National Science Foundation under grant
DMS-9704639.

\begin{appendix}
\section{Appendix: Kneading sequences and polynomials for the gap tent
map}

If the trajectory of the critical point is periodic
the topological entropy of the gap tent map $f_{\varepsilon}$ can be
expressed as a logarithm of largest root of some polynomial,
with all coefficients equal to $\pm 1$. Even though  this fact
follows from the kneading theory of Milnor and Thurston \cite{MT77}, we
give here a brief derivation of this result and introduce the
polynomials and notation used in this paper.

 Our reasoning is based on the fact that two conjugate
 maps share the same topological entropy \cite{Ott94}.
For the map $f_{\varepsilon}$
 (or for other unimodal maps) it is sufficient
to find such a value of the slope $s$
of the tent map $f_s(x):=s(1/2-|x-1/2|)$, that the kneading sequences
are
identical. Then the entropy of the analyzed map is equal to ln$s$ - the
entropy of $f_s$ \cite{MeloStrein}. For the simple orbit {\bf CR} of
length $2$ the condition $f_s^2(1/2)=1/2$ leads to following
equation $s(1-s/2)=1/2$. It can be rewritten as
$(s-1)P_2(s)=0$, with $P_2(s)=(-s+1)$ represented by $[- +]$.

Proceeding inductively, we assume that a sequence {\bf Q} of
length
$L$ corresponds to the polynomial $P_L=[c_{L-1},\dots,c_1, c_0]$.
Extending the kneading sequence by one symbol, {\bf Q} $\to$ {\bf QX},
the descendent polynomial $P_{L+1}(s)$ reads

\begin{equation}
P_{L+1}(s)  =  \left\{ \begin{array}{lll}
 s P_L(s) +1,~&~{\rm{for}}~&~{\bf X=L} \nonumber \\
-s P_L(s) +1,~&~{\rm{for}}~&~{\bf X=R}. \\
\end{array}
\right.
\label{indu}
\end{equation}
Therefore, every coefficient of any polynomial
is equal to $\pm1$. Since multiplication of all coefficients of a given
polynomial by $-1$ does not influence its roots, we can
arbitrarily define the leading coefficient, $c_{L-1}$, to
be $+1$. This corresponds to the initial symbol {\bf C} (strictly
speaking
it should be {\bf L} for left end of the plateau and {\bf R} for
the right one).
 The next sign of the polynomial is determined by the next symbol of the
kneading
sequence: when the symbol is an {\bf L}, the sign
is the same as the previous sign, while the sign changes when the symbol
is an
{\bf R}. More precisely,
\begin{equation}\label{coeffs0}
c_{L-1}=+1\mbox{, and }c_j=\prod_{i=j}^{L-2} g_i\mbox{, for
}j=0,\dots,L-2,
\end{equation}
where $g_i(${\bf L}$)=+1$ and $g_i(${\bf R}$)=-1$ \cite{MT77}.
See several examples collected in Table 1.

Any entropy plateau occurring for $\varepsilon \in B_{mk}$
may be related with a  concrete periodic orbit. Consider,  for example,
$\varepsilon_{m1}=1/(2^{m+3}+1)$, which corresponds to
left edges of plateaus  $B_{m1}$ defined in Eq. (\ref{cons}).
The orbit starting at $x=2\varepsilon_{m1}$
is periodic with the length $L=m+3$,
 and its kneading sequence reads {\bf CRLL}$\dots${\bf L}.
 The corresponding polynomial $[+ - - \cdots - -]$ can be found
independently by the companion matrix \cite{matrix} which
 is equal to the topological transition matrix of the system
$f_{\varepsilon_L}$.
The largest eigenvalues of the topological transition matrices can be
used
 to find the topological entropy - see eg. \cite{DGP78,HB89,GB91}.

Let us order entropy plateaus $B_{m,k}$, corresponding to
$m$-th preimages of the gap $A_{10}$, according to decreasing entropy.
When we increase the gap width $\varepsilon$, we decrease the critical
point
$x_c=(1-\varepsilon)/2$. Since the real line is ordered
monotonically with the order of kneading sequences \cite{MT77,De89} (and
polynomials),
the $k$-th plateau $B_{m,k}$
corresponds to the $k$-th periodic orbit of the length $L=m+3$
(ordered according to decreasing entropy).
In other words, the periodic orbit represented by the polynomial
$[c_{L-1},c_{L-2},\dots,c_1,c_0]$ is $k$-th in the family of
orbits of length $L$, where
\begin{equation}\label{coeffs}
k=1+ \sum_{j=0}^{L-2} 2^j (c_j+1)/2.
\end{equation}
Thus, this orbit corresponds to the plateau $B_{L-3,k}$, which
occurs for the gap sizes, $\varepsilon$, determined by Eq. (\ref{cons})
-
see Table 1.

\section{Appendix: Gap--Tent Map and Iterated Function System}

We describe here a technique of generating invariant
measures for dynamical systems via appropriate IFS as applied in
Refs. \cite{BDEG88,GB95,SKZ98}.  In chaos, the initial condition is
chosen randomly, albeit in a small
ball, (e.g. machine precision in {\it double} is typically a ball of
radius
$\approx 10^{-16}$), and the sensitive dependence to initial conditions,
of
the nonetheless deterministic dynamical system, amplifies this
randomness.
The deterministic chaos problem, can be traded for an appropriately
chosen
truly stochastic process, which evolves (supposedly exact) initial
conditions
by a random dynamical system, whose randomness mimics the chaos.

Barnsley's Iterated Function Systems (IFS) \cite{Barnsley} are an
appropriate
formalism by which we may accurately exchange the deterministic problem
for
the right stochastic problem.  In simplest form, an IFS of the first kind
 involves
an iteration $x_{n+1}=F_i(x_n)$, where the function actually used at
each
step is chosen randomly with place dependent probabilities
$\{p_i(x) \}_{i=1}^k$,
 $\sum p_i(x)=1$,
amongst  $k$ possible functions $\{F_i(x)\}_{i=1}^k$.

For the tent gap-map model, we define an IFS consisting of two
functions with place dependent probabilities:
  $\{ X=[2\varepsilon, 1-\varepsilon], ~
   F_1(x)=x/2, F_2(x)=1-x/2;$ $p_1= 0$ for $x<4\varepsilon$ and
   $p_1=w$ for $x \ge 4\varepsilon;$ $p_2=1-p_1 \}$,
  where the relative weight $w$
is a free parameter.  Since there exist points
transformed by one function  with probability one
($p_2=1$ for $x\in [2\varepsilon, 4\varepsilon)$), the standard
assumptions \cite{Barnsley,BDEG88} sufficient to prove existence of a
unique attracting invariant measure are not fulfilled for this IFS.

Nonetheless we conjecture:

 a)  - for every value of $w\in (0,1)$ there exist an attracting
 invariant measure $\nu_w$ of the IFS and it is localized on the same
 support $S$ as the measure $\mu_{SRB}$.

 b) - for every value of $\varepsilon$, there exist
      $w=w(\varepsilon)$
      such that the induced invariant measure of the IFS, $\nu_w$, and
      the SRB measure of the tent map with the gap, are equal:
      $\mu_{SRB} =\nu_w$.

 c) - the spectrum of entropies $K_q$ and
      the generalized dimensions $D_q$ for
      IFS$(\varepsilon, w(\varepsilon))$  and the tent map with a gap
      $f_{\varepsilon}$ are identical for any fixed value of
      $\varepsilon$.

\smallskip
Let us consider the simplest case with the gap of the width
 $\varepsilon=1/7$ for which $M=\gamma$ (compare to Example 1). Since
the interval
$(4/7,6/7)$ of the mass $M$ is transformed by this IFS, with probability
$w$, into the interval $(2/7,3/7)$ of mass $1-M$,
the relative weight $w$ is equal to $(1-M)/M=(1-\gamma)/\gamma=\gamma$.
 More
generally, for $\varepsilon \in B_{m1}$ the above relation is fulfilled
for $w=1/\lambda_{m1}$, so in the limit of no gap $\varepsilon \to 0$
 one has  $lim_{m\to \infty} \lambda_{m1}=2$ and
 the IFS becomes symmetric ( $w=p_1=p_2=1/2$).

\end{appendix}

%\newpage

{}

\medskip
 {\bf Figure captions}
\smallskip
\vskip 0.1cm
{\bf Figure 1} Tent map with an $\varepsilon$-gap represented
 by the central dark strip. Its preimages of the first and second
order
are represented by narrower vertical strips. The support of the
invariant measure can be divided into two parts: $E_2$, for which
each point has two preimages and $E_1$ for which there exist only one
preimage.

\smallskip
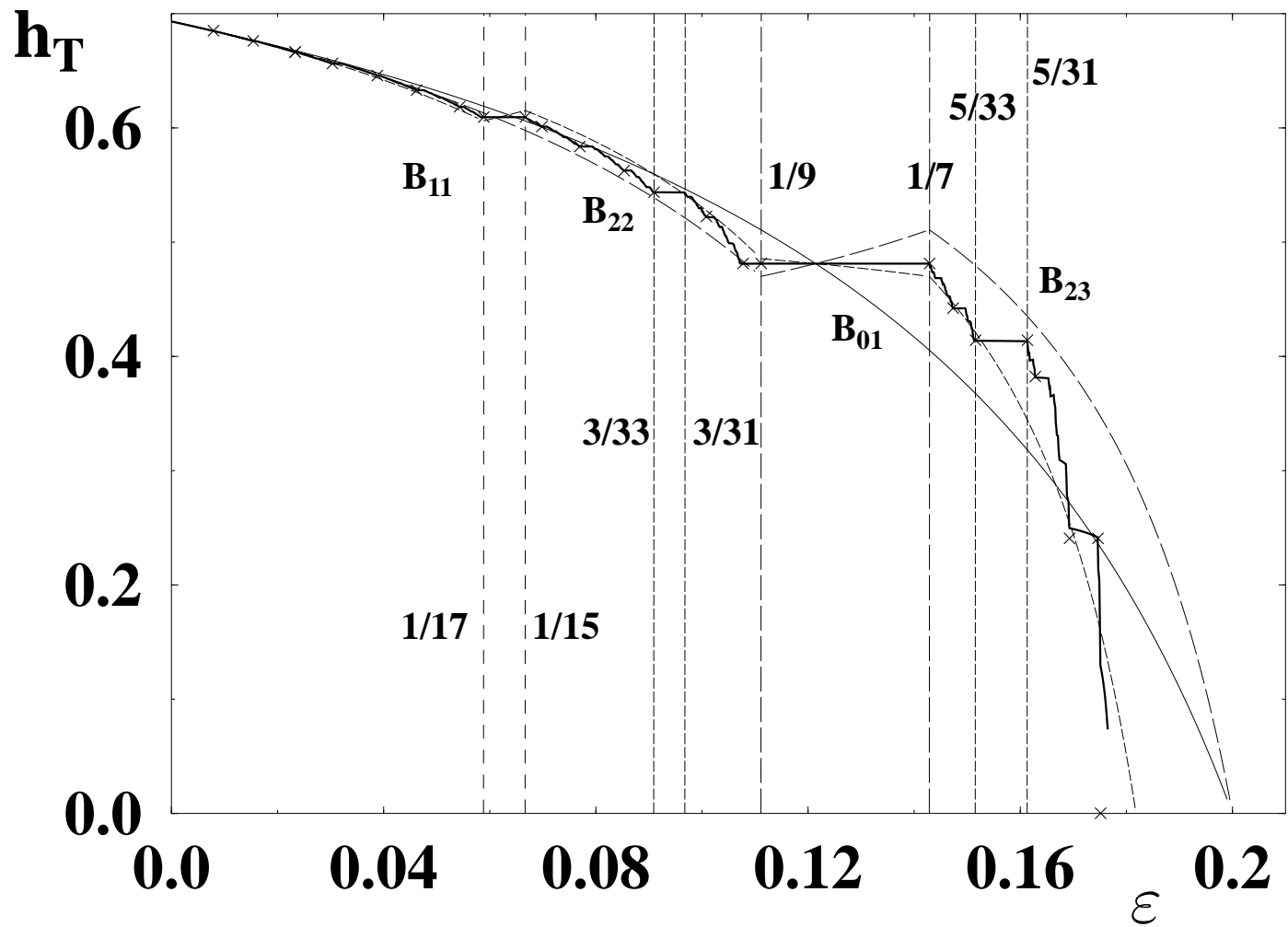
{\bf Figure 2} Topological entropy $h_T$ of the tent map
 as a function of the gap width $\varepsilon$ obtained via
 Eq. (\ref{entr}) (thick line). Crosses represent
 points at the edges of the entropy plateaus computed from roots
 of polynomials (\ref{poln}).
Position of the main plateaus are labeled according to the relation
(\ref{cons}).
Narrow  solid, dashed and dotted lines
represent continuous approximations of $h_0(\varepsilon)$,
$h_1(\varepsilon)$, and $h_2(\varepsilon)$ respectively,
(Eqs.~\ref{a0} - \ref{a2})
 of the zero-th, first, and second order, respectively.

\smallskip
{\bf Figure 3} Sketch of the dependence of topological entropy
$h_T$ on the gap width $\varepsilon$  in the vicinity of the critical
point $e_*=\epsilon_*$.

\medskip

{\bf Table 1.}
    Topological entropy at plateaus and the corresponding periodic
orbits for the tent map $2x$ with symmetric
$\varepsilon$-gap for $x \in [(1-\varepsilon)/2,(1+\varepsilon)/2]$.
Subsequent columns contain respectively : length of the orbit $L$, root
$\lambda$,
topological entropy equal to ln$(\lambda)$, polynomial, number
$k-1$ labeling the $L-3$-th preimages of the critical point,
 kneading sequence,  both edges of the plateau $\varepsilon_-$ and
$\varepsilon_+$. The symbol $[+ -  -]$ represents  the polynomial
$x^2-x-1=0$, which root gives the
golden mean. The letter $E$ denotes an descendent orbit
forming an extension of the plateau
 related  to the two times shorter ancestor orbit,
 $N$ denotes an non-admissible orbit, which plateau is  entirely
shadowed
by the ancestor plateau. The symbol $D$ represents descendents orbits
obtained by the
renormalization operation $W_1$: two times longer period corresponds to
  two times smaller topological entropy.

\vskip 1.0cm
\begin{tabular}{|r|c|c|l|c|l|c|c|}
\hline
\hline
 Length  &  $\lambda$  & Entropy   & Polynomial & k-1 & Kneading Seq. &
 $\varepsilon_-$ & $\varepsilon_+$  \\
\hline
\hline
 7 & ~1.9835828~ & ~0.6849047~ &~[+ - - - - - -]~& 0 &~CRLLLLL &
 0.00775 & 0.00787 \\
  6 & 1.9659482 & 0.6759747 &~[+ - - - - -]  &  0 &~CRLLLL &
0.01538 & 0.01587 \\
 7 & 1.9468563 & 0.6662159 &~[+ - - - - - + ] &  1 &~CRLLLLR &
0.02326 &  0.02362 \\
 5 & 1.9275620 & 0.6562560 &~[+ - - - - ]  & 0 &~CRLLL & 0.03030 &
0.03226 \\
 7 & 1.9073421 & 0.6457107 &~[+ - - - - + -] &  2 &~CRLLLRR &
0.03876 &  0.03937 \\
 6 & 1.8832035 & 0.6329743 &~[+ - - - - +] & 1 &~CRLLLR & 0.04615
 & 0.04762 \\
 7 & 1.8558860 & 0.6183622 &~[+ - - - - + +] & 3 &~CRLLLRL
 & 0.05426 & 0.05512 \\
E 8 & 1.8392868 & 0.6093779 &~[+ - - - - + + +] &  7 &~CRLLLRLL &
    0.05837  & 0.05882 \\
 4 & 1.8392868 & 0.6093779  &~[+ - - -] &  0  &~CRLL & 0.05882 &
   0.06667 \\
N 8 & 1.8392868 & 0.6093779 &~[+ - - - + - - -] &  8 &~CRLLRRLL &
    0.06615 & 0.06667 \\
 7 & 1.8239445 & 0.6010015 &~[+ - - - - -]
   &   4 &~CRLLRRL &  0.06977 & 0.07087 \\
 6 & 1.7924024 & 0.5835568 &~[+ - - - + -] &  2  &~CRLLRR &   0.07692
&  0.07937 \\
 7& 1.7548777 & 0.5623992  &~[+ - - - + - +] & 5 &~CRLLRRR & 0.08527
 & 0.08661 \\
 5 & 1.7220838 & 0.5435351 &~[+ - - - + ] & 1 &~CRLLR &
  0.09091 & 0.09677 \\
 7 & 1.6859262 & 0.5223151 &~[+ - - - + + -] & 6 &~CRLLRLR &
 0.10078 & 0.10236 \\
E 6 & 1.6180340 & 0.4812118 &~[+ - - - + + ] &  3 &~CRLLRL &
0.10769 & 0.11111 \\
 3 & 1.6180340 & 0.4812118  &~[+ - -]   &   0 &~CRL & 0.11111 &
 0.14286 \\
N 6 & 1.6180340 & 0.4812118  &~[+ - - + - -] & 4 &~CRLRRL & 0.13846 &
   0.14286 \\
 7 & 1.5560302 & 0.4421378  &~[+ - - + - - +] & 9  &~CRLRRLR &
  0.14729 & 0.14961 \\
 5& 1.5128764 & 0.4140127 &~[+ - - + -] &  2 &~CRLRR & 0.15152
 & 0.16129 \\
 7 &1.4655712 & 0.3822451 &~[+ - - + - + -]  & 10 &~CRLRRRR &
 0.16279 & 0.16535 \\
D 8 & 1.3562031 & 0.3046889 &~[+ - - + - + - +] &  21 &~CRLRRRRR &
    0.16732 & 0.16863 \\
D 6 & 1.2720197 & 0.2406059 &~[+ - - + - +] &  5 &~CRLRRR &
    0.16923 & 0.17460 \\
\hline
%  ZERO entropy plateau
DE 8 & 1.0  &      0.0   &~[+ - - + - + + -]~& 22 &~CRLRRRLR &
   0.17510 & 0.17647 \\
DE 4 & 1.0  &      0.0   &~[+ - - +] &     1 &~CRLR &     0.17647 &
0.20000 \\
2 & 1.0  &      0.0   &~[+ -]    &     0 &~CR   &   0.20000 &
0.33333 \\
\hline
\hline
\end{tabular}

\vskip 1.0cm
{\bf Table 2.}
Number $I_L$ of
periodic orbits of length $L$ creating a plateau in the entropy
dependence $K(\varepsilon)$ for
$\varepsilon \in \Delta_0=(0,1/6)$.
Total length of these plateaus equals $T_L$. Cumulative number of
plateaus $I_{tot}=\sum_{k=3}^L I_L$, while
$W_L=1/6 - \sum_{k=3}^L T_L$
represents the total volume of the parameter space {\sl not included} in
the plateaus.

\vskip 1.0cm
\begin{tabular}{|c|r|r|c|c|}
\hline
\hline
~$L$~~&~~~$I_L$~~~&~~~$I_{tot}$~~~&~~~$T_L$~~~&~~~$W_L$~~~\\
\hline
\hline
  3 &     1  &     1 &  0.03175 &  0.13492   \\
  4 &     1  &     2 &  0.00784 &  0.12708   \\
  5 &     3  &     5 &  0.01759 &  0.10948   \\
  6 &     3  &     8 &  0.00781 &  0.10167   \\
  7 &     9  &    17 &  0.01086 &  0.09080   \\
  8 &    13  &    30 &  0.00659 &  0.08421   \\
  9 &    28  &    58 &  0.00720 &  0.07701   \\
 10 &    45  &   103 &  0.00522 &  0.07179   \\
 11 &    93  &   196 &  0.00528 &  0.06651   \\
 12 &   161  &   357 &  0.00412 &  0.06238   \\
 13 &   315  &   672 &  0.00396 &  0.05842   \\
 14 &   567  &  1239 &  0.00330 &  0.05512   \\
 15 &  1091  &  2330 &  0.00307 &  0.05205   \\
 16 &  2018  &  4348 &  0.00267 &  0.04938   \\
 17 &  3855  &  8203 &  0.00247 &  0.04692   \\
\hline
\hline
\end{tabular}

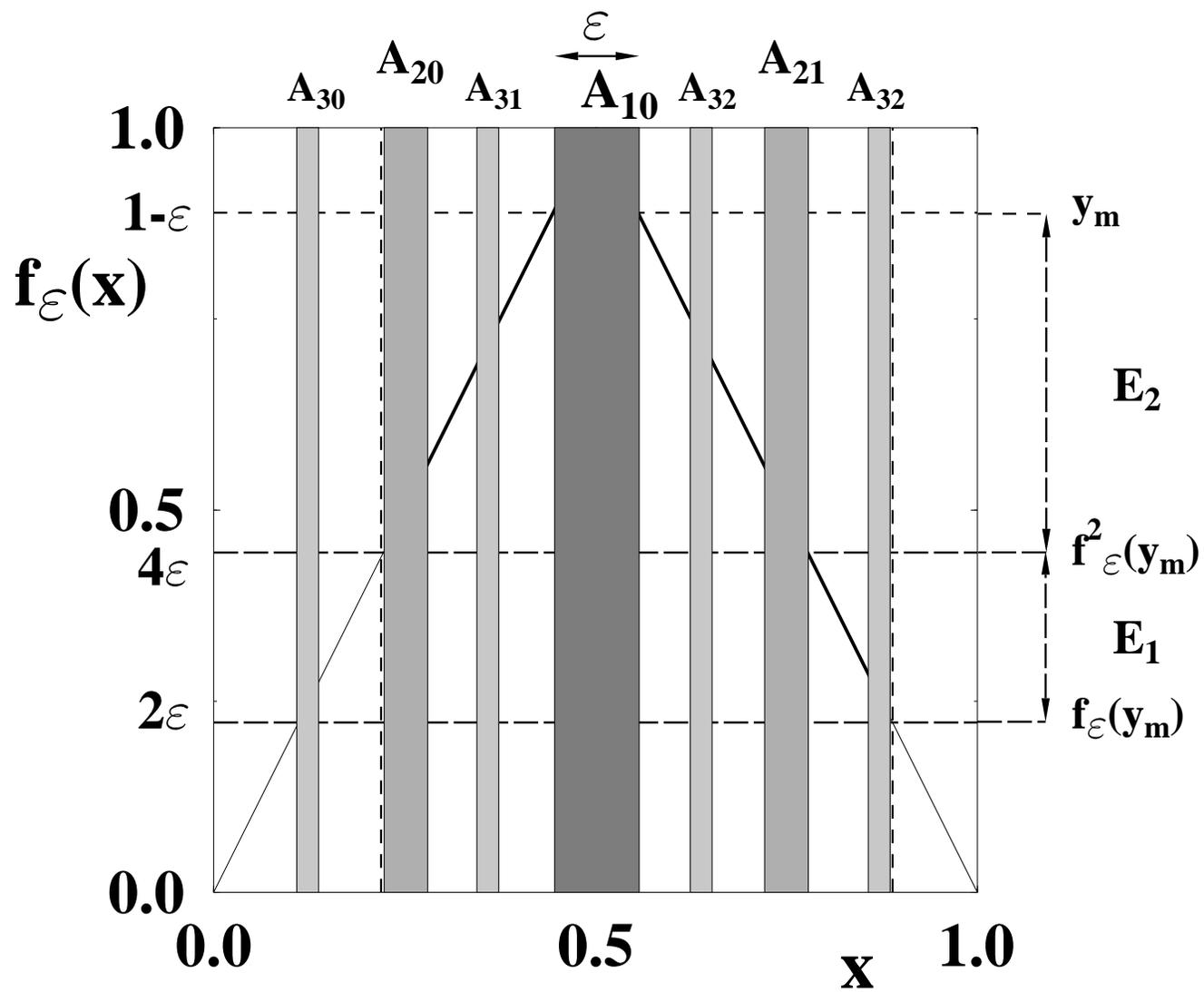
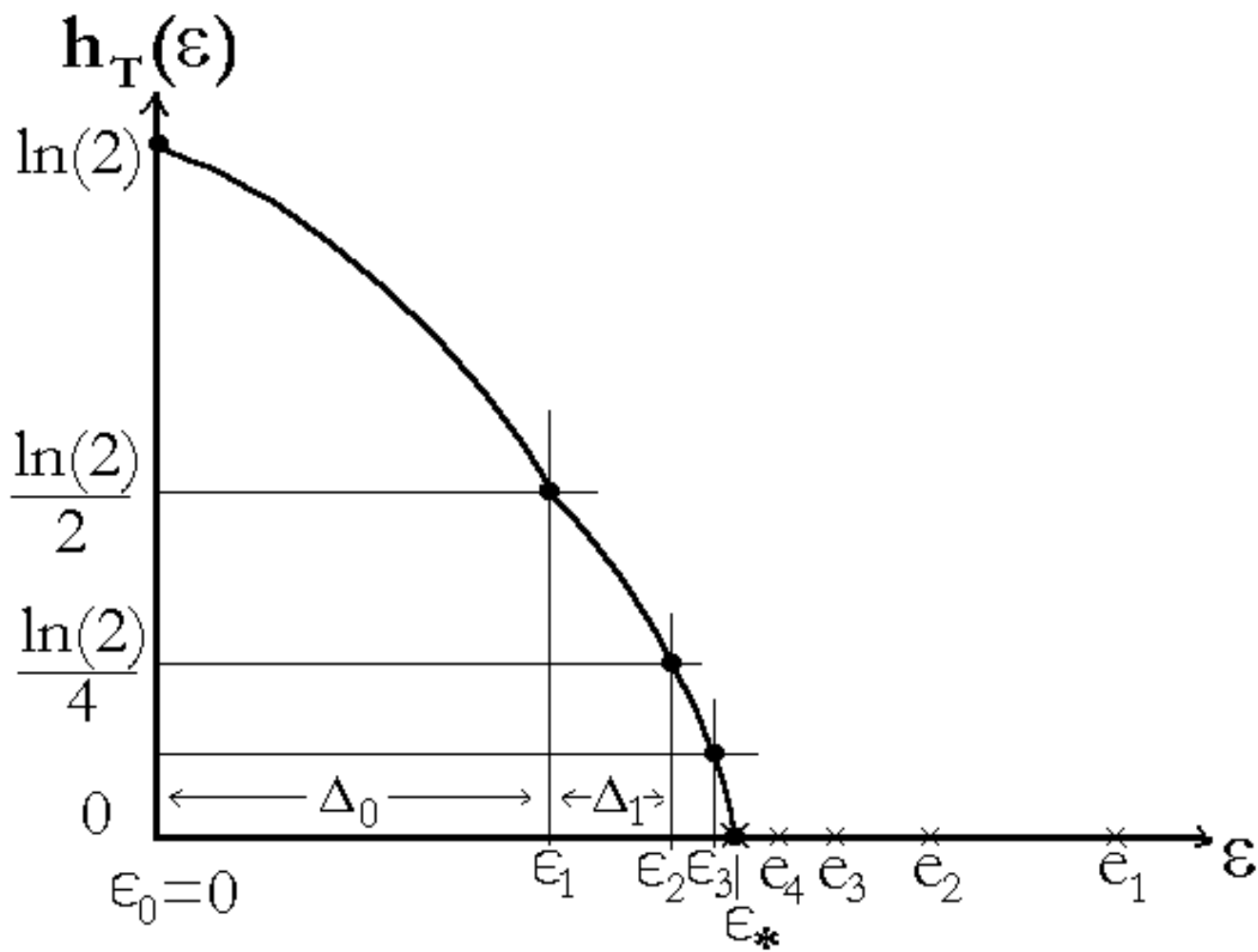
\end{document}